\documentclass[aps,pra,amsmath,amssymb,preprint]{revtex4-1}

\usepackage[dvipsnames]{xcolor}
\definecolor{myblue}{HTML}{1F77B4}
\definecolor{myorange}{HTML}{FF7F0E}
\usepackage{hyperref}
\hypersetup{
	colorlinks,
	linkcolor={red!50!black},
	citecolor={red!70!black},
	urlcolor={red!80!black}
}
\usepackage[capitalize]{cleveref}
\usepackage{graphicx}

\newcommand{\Pu}{P{\mathrel{\uparrow}}}
\newcommand{\Pd}{P{\mathrel{\downarrow}}}
\newcommand{\Qu}{Q{\mathrel{\uparrow}}}
\newcommand{\Qd}{Q{\mathrel{\downarrow}}}
\newcommand{\Ru}{R{\mathrel{\uparrow}}}
\newcommand{\Rd}{R{\mathrel{\downarrow}}}
\newcommand{\Su}{S{\mathrel{\uparrow}}}
\newcommand{\Sd}{S{\mathrel{\downarrow}}}

\newcommand{\sm}{Supplemental Material}
\newcommand{\foreign}[1]{\textit{#1}}

\newcommand{\abs}[1]{\lvert#1\rvert}
\newcommand{\norm}[1]{\lVert#1\rVert}

\newcommand{\ket}[1]{\lvert #1\rangle}

\newcommand{\braket}[2]{\langle #1 \!\mid\! #2 \rangle}
\newcommand{\presuperscript}[2]{\; {\vphantom{#1}}^{#2}\!{#1}}

\usepackage{pdfpages}

\makeatletter
\AtBeginDocument{\let\LS@rot\@undefined}
\makeatother

\begin{document}

\title{Closed-Form Expressions for Unitaries of Spin-Adapted Fermionic Operators}

\author{Ilias Magoulas}
\email{ilias.magoulas@emory.edu}
\author{Francesco A. Evangelista}
\affiliation{Department of Chemistry and Cherry Emerson Center for Scientific Computation, Emory University, Atlanta, Georgia 30322, USA}

\date{\today}

\begin{abstract}
One of the open challenges in quantum computing simulations of problems of chemical interest is the proper enforcement of spin symmetry.
Efficient quantum circuits implementing unitaries generated by spin-adapted operators remain elusive, while na\"{\i}ve Trotterization schemes break spin symmetry.
In this work, we analyze the mathematical structure of spin-adapted operators and derive closed-form expressions for unitaries generated by singlet spin-adapted generalized single and double excitations.
These results represent significant progress toward the economical enforcement of spin symmetry in quantum simulations.
\end{abstract}

\maketitle

\section{Introduction}

Accurate and predictive classical simulations of quantum many-body systems quickly become intractable due to the combinatorial growth of the required computational resources with the system size.
In the philosophy of ``fight fire with fire,'' in 1982, Feynman proposed the use of a machine built from quantum mechanical elements, called qubits, to imitate any desired quantum system \cite{Feynman.1982.10.1007/BF02650179}.
The power of such quantum devices lies in their ability to represent highly entangled quantum many-body states with a number of qubits that scales proportionately with the system size. 
Manipulations of qubits are achieved via quantum gates, which, mathematically, are represented by unitary operators.

Despite the various technological advances that have taken place in the more than 40 years following Feynman's vision \cite{Preskill.2021.2106.10522v2}, the currently available quantum hardware is quite ``noisy'' \cite{Preskill.2018.10.22331/q-2018-08-06-79}.
Noise in the context of quantum computing refers to errors that can arise during quantum operations that can greatly degrade the performance and accuracy of quantum simulations.
This issue could be addressed using quantum error correction to create robust, logical qubits \cite{Campbell.2017.10.1038/nature23460}.
Such a setting would enable the routine application of pure quantum algorithms, which typically require deep circuits.
One such example is quantum phase estimation \cite{Kitaev.1995.quant-ph/9511026}, the ``holy grail'' for the simulation of quantum many-body systems. 
This algorithm can efficiently extract the eigenvalues of unitary matrices with a potential exponential speedup compared to classical methods \cite{Lee.2023.10.1038/s41467-023-37587-6}.
However, the construction of a single logical qubit requires thousands of physical ones \cite{Campbell.2017.10.1038/nature23460} (see, however,  \cite{Bravyi.2024.10.1038/s41586-024-07107-7}).
At the same time, scaling-up the size of quantum devices is very challenging, being called ``the experimenter's nightmare'' \cite{Haroche.1996.10.1063/1.881512}.
As a result, fault-tolerant computing is currently out of reach, although promising results indicate that we are entering the early fault-tolerant era in which error correction is realized only partially \cite{Katabarwa.2024.10.1103/PRXQuantum.5.020101,Zhang.2025.2502.02139}. 

To tackle the noise challenge from an algorithmic perspective, hybrid quantum--classical schemes have been introduced 
\cite{Endo.2021.10.7566/JPSJ.90.032001,Callison.2022.10.1103/PhysRevA.106.010101}.
These rely on a quantum device to access information that would be computationally demanding, or even intractable, classically.
A classical computer is then used to process the information and optimize the various parameters until an optimal solution is found.
Such approaches require shorter and shallower circuits than pure quantum algorithms.
This results in more robust quantum simulations as the errors associated with the experimental realization of quantum gates as well as decoherence errors are reduced. 

For the purposes of this work, we focus on ansatz-dependent hybrid schemes, where a trial state $\ket{\Psi}$ of the desired many-body system is constructed in terms of a unitary parameterization of an initial, typically separable, state $\ket{\Phi}$,
\begin{equation}
	\ket{\Psi(\boldsymbol{\theta})} = U(\boldsymbol{\theta}) \ket{\Phi} \equiv \prod_i U_i(\theta_i) \ket{\Phi}.
\end{equation}
Depending on the form of the unitaries $U_i$, there exist two popular families of ans\"{a}tze.
In the first one, the $U_i$ gates represent the elementary native gates available on a given device \cite{Kandala.2017.10.1038/nature23879}.
These types of ans\"{a}tze are adapted to hardware limitations and, thus, their experimental implementation is natural.
Although hardware-efficient ans\"{a}tze typically require shallower circuits, they often lack clear physical interpretation, are not guaranteed to converge to the exact solution, and typically break the symmetries of the chemical system. 
In the second category, one relies on chemically motivated ans\"{a}tze derived from the unitary extension \cite{Kutzelnigg.1977.10.1007/978-1-4757-0887-5_5,Kutzelnigg.1982.10.1063/1.444231,
	Kutzelnigg.1983.10.1063/1.446313,Kutzelnigg.1984.10.1063/1.446736,Bartlett.1989.10.1016/S0009-2614(89)87372-5,
	Szalay.1995.10.1063/1.469641,Taube.2006.10.1002/qua.21198,Cooper.2010.10.1063/1.3520564,
	Evangelista.2011.10.1063/1.3598471,Harsha.2018.10.1063/1.5011033,Filip.2020.10.1063/5.0026141,
	Freericks.2022.10.3390/sym14030494,Anand.2022.10.1039/d1cs00932j}
of coupled-cluster (CC) theory \cite{Coester.1958.10.1016/0029-5582(58)90280-3,
	Coester.1960.10.1016/0029-5582(60)90140-1,Cizek.1966.10.1063/1.1727484,Cizek.1969.10.1002/9780470143599.ch2,
	Cizek.1971.10.1002/qua.560050402,Paldus.1972.10.1103/PhysRevA.5.50} (UCC).
In contrast to hardware-efficient schemes, UCC is physically motivated, it systematically converges to the exact solution, and the enforcement of symmetries is easier.
To that end, in this work we focus on UCC-type ans\"{a}tze.
The parameters $\boldsymbol{\theta} = (\theta_1, \theta_2, \ldots)$ are 
optimized classically utilizing information extracted from measurements performed on a quantum device.
Depending on the type of information, we have approaches such as the variational (VQE) 
\cite{Peruzzo.2014.10.1038/ncomms5213,McClean.2016.10.1088/1367-2630/18/2/023023,
	Cerezo.2021.10.1038/s42254-021-00348-9,Tilly.2022.10.1016/j.physrep.2022.08.003,Fedorov.2022.10.1186/s41313-021-00032-6},
projective (PQE) \cite{Stair.2021.10.1103/PRXQuantum.2.030301}, and contracted (CQE) \cite{Smart.2021.10.1103/PhysRevLett.126.070504} quantum eigensolvers.
Although hybrid approaches are still limited by the classical computational resources, they enable the use of powerful methods, such as UCC, that are intractable on classical machines.

Drawing inspiration from selected/adaptive computational chemistry approaches 
\cite{Whitten.1969.10.1063/1.1671985,Bender.1969.10.1103/PhysRev.183.23,Huron.1973.10.1063/1.1679199,
	Buenker.1974.10.1007/BF02394557},
hybrid schemes relying on iteratively constructed ans\"{a}tze have been proposed to further reduce the required quantum resources.
A crucial concept in these algorithms is that of the operator pool, \foreign{i.e.}, the set of operators used to grow the ansatz.
Methods in this category include the adaptive derivative-assembled pseudo-Trotter ansatz VQE (ADAPT-VQE) \cite{Grimsley.2019.10.1038/s41467-019-10988-2}, iterative qubit coupled cluster (iQCC) \cite{Ryabinkin.2020.10.1021/acs.jctc.9b01084}, and selected PQE \cite{Stair.2021.10.1103/PRXQuantum.2.030301}.
For example, in ADAPT-VQE the ansatz is built by selecting and adding operators from a predefined operator pool based on the energy gradient with respect to each operator, variationally optimizing all parameters at the end of each macro-iteration.
If ans\"{a}tze generated from a given operator pool have the ability to cover/explore the entire $N$-electron Hilbert space, then the pool is termed universal.
In principle, an adaptive approach relying on a universal operator pool will converge to the global energy minimum, \foreign{i.e.}, the ground-state full configuration interaction (FCI) solution.
It is generally accepted that an operator pool containing all possible one-body, two-body, \ldots, $N$-body 
particle--hole excitation operators, where $N$ is the number of correlated electrons, is universal. Although the 
generalized singles and doubles (GSD) \cite{Nooijen.2000.10.1103/PhysRevLett.84.2108,Nakatsuji.2000.10.1063/1.1287275} 
pool is not \foreign{per se} universal, it can be made so if each operator is 
allowed to be added to the ansatz multiple times \cite{Evangelista.2019.10.1063/1.5133059}, albeit with a different 
optimization parameter.

To further optimize the performance of adaptive algorithms, one can construct symmetry-preserving operator pools.
Typical examples of symmetries encountered in molecular systems include point group, particle number ($N$), $z$-component of total spin ($S_z$), and the square of the total spin ($S^2$), assuming a non-relativistic setting.
Although the enforcement of the former three symmetries is rather straightforward for UCC-type operator pools, the efficient construction of spin-adapted quantum circuits remains elusive.
The main problem stems from the fact that na\"{\i}ve Trotterization schemes of exponentials of spin-adapted operators break spin symmetry and are prone to variational collapse to a state with the wrong symmetry \cite{Tsuchimochi.2020.10.1103/PhysRevResearch.2.043142}.
Various solutions have been proposed over the years for realizing spin-adapted quantum circuits.
Notable examples include the symmetry-preserving state preparation circuits \cite{Gard.2020.10.1038/s41534-019-0240-1} and approaches relying on an operator pool of spin-adapted singles and perfect pairing doubles (saGSpD) \cite{Anselmetti.2021.10.1088/1367-2630/ac2cb3,Burton.2023.10.1038/s41534-023-00744-2,Burton.2024.10.1103/PhysRevResearch.6.023300}.
At this point, it is also worth mentioning approaches that enforce spin symmetry without explicitly constructing spin-adapted quantum circuits.
Examples in this category include Hamiltonian penalty \cite{Ryabinkin.2019.10.1021/acs.jctc.8b00943} and projection \cite{Tsuchimochi.2020.10.1103/PhysRevResearch.2.043142,Tsuchimochi.2022.10.1103/PhysRevResearch.4.033100} techniques (see, also, \cite{Selvarajan.2022.10.3390/sym14030457} for penalty-dependent approach to construct symmetry-preserving quantum circuits).
Despite the usefulness of the aforementioned schemes, they come with their own issues.
The symmetry-preserving state preparation circuits enforce spin symmetry by constructing a gate that acts on all qubits in the system, essentially constructing the FCI state within a hyperspherical parameterization.
Although the saGSpD-based approaches are spin-adapted, universality can only be attained by breaking point group symmetry.
Spin-penalty techniques introduce a number of terms in the Hamiltonian that scale as $N^8$, where $N$ is the number of spinorbitals, dramatically increasing the number of Pauli measurements (however, see \cite{Li.2024.10.1038/s43588-024-00730-4} for a more efficient formulation based on states with higher $S_z$ values).
Restoring spin symmetry via projection requires an ancilla qubit, additional two-qubit gates, and an increase in the number of Pauli measurements by at least a factor of 2.
Parallel to these developments, it is worth mentioning efforts toward realizing total-spin eigenstates on quantum devices \cite{Sugisaki.2016.10.1021/acs.jpca.6b04932,Sugisaki.2019.10.1016/j.cpletx.2018.100002,Carbone.2022.10.3390/sym14030624}.

In this work, we explore the mathematical intricacies of unitaries of spin-adapted operators.
In particular, we are focusing on an operator pool comprised of singlet spin-adapted single and double excitations.
We provide evidence why their efficient quantum-circuit implementation is difficult, showcasing the failure of finite Trotter--Suzuki decompositions.
As a result of our mathematical explorations, we present closed-form expressions of unitaries of singlet spin-adapted singles and doubles.
Our findings offer insights into the intrinsic complexity of spin-adapted quantum circuits and open up new possibilities for constructing symmetry-preserving variational algorithms.

While finalizing this manuscript, we became aware of concurrent work by Kjellgren \foreign{et al.} \cite{Erik}, who independently derived closed-form expressions for the matrix exponential of the anti-Hermitian form of singlet spin-adapted double excitations, expressed in terms of powers of the pertinent generator.
Although our two studies differ methodologically (our approach derives closed-form formulas first in spinorbital form and subsequently translates them into expressions involving generator powers, whereas Kjellgren \foreign{et al.} directly establish recursion relations between generator powers) the resulting expressions are identical. Indeed, our \cref{eq:sa_unitary_closed3} and Eqs. [S9] and [S10] in the {\sm} exactly match their Eqs. [41], [43], and [45].
All closed-form expressions presented herein, both in spinorbital and power forms, have been thoroughly verified numerically.

\section{Symmetries and Quantum Simulations}

\begin{figure}[h!]
	\centering
	\includegraphics[width=3.375in]{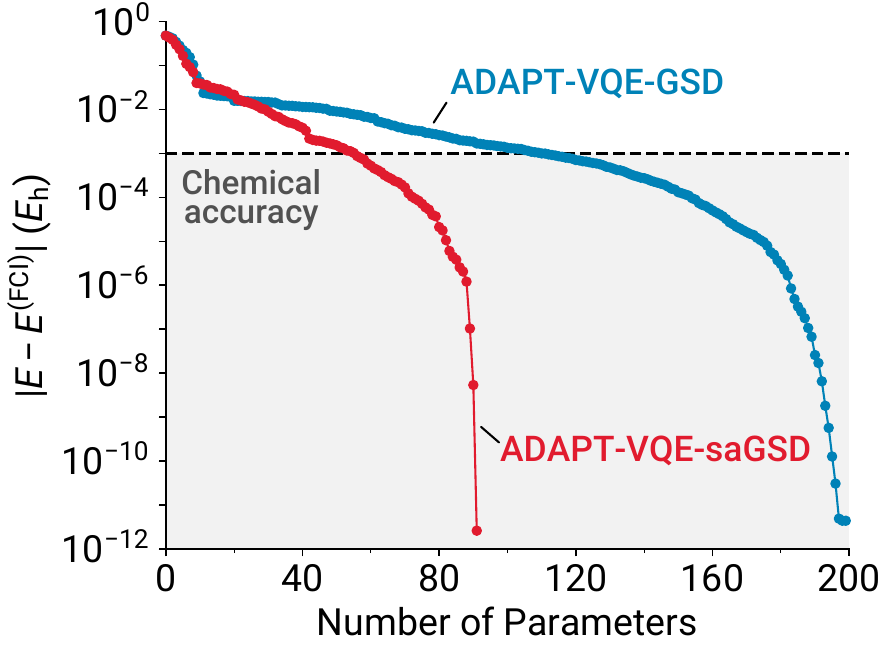}
	\caption{Errors relative to FCI characterizing the ADAPT-VQE-GSD and ADAPT-VQE-saGSD simulations of the 
		H$_6$/STO-6G linear chain with $R_\text{H--H} = 2.0\, \text{\AA}$.}
	\label{GSD_vs_asGSD}
\end{figure}

Similar to classical electronic structure approaches, using a symmetry-adapted operator pool has various advantages.
To begin with, the enforcement of symmetries substantially shrinks the size of the operator pool and the targeted Hilbert space of the system of interest.
For example, in the linear H$_6$ chain as described by the STO-6G basis \cite{Hehre.1969.10.1063/1.1672392}, the GSD pool contains 1551, 870, 420, and 312 operators when the ($N$), ($N$, $S_z$), ($N$, $S_z$, point group), and ($N$, $S_z$, point group, $S^2$) symmetries are enforced, respectively.
Furthermore, the dimensions of the symmetry-adapted Hilbert spaces are 924 ($N = 6$), 400 ($N = 6$, $S_z = 0$), 200 ($N = 6$, $S_z = 0$, $A_g$ [$D_{2h}$]), and 92 ($N = 6$, $S_z = 0$, $A_g$ [$D_{2h}$], $S^2 = 0$).
In \cref{GSD_vs_asGSD}, we compare the convergence to the exact, FCI ground-state energy of H$_6$/STO-6G, $R_\text{H--H} = 2.0\, \text{\AA}$, for ADAPT-VQE numerical simulations using the standard GSD pool and its singlet spin-adapted variant (saGSD).
Although both pools respect the $N$, $S_z$, and point group symmetries, only the saGSD pool is singlet spin-adapted, whereas the GSD pool is not.
As shown in \cref{GSD_vs_asGSD}, both ADAPT-VQE-GSD and ADAPT-VQE-saGSD ultimately reproduce the FCI result within a few picohartree.
However, ADAPT-VQE-saGSD requires only 91 parameters, compared to 199 for its GSD counterpart, mirroring the differences in the dimensions of their respective symmetry-adapted Hilbert spaces.
As might have been anticipated, ADAPT-VQE-saGSD reaches chemical accuracy much faster, requiring less than half the number of parameters of ADAPT-VQE-GSD.

\begin{figure}[!h]
	\centering
	\includegraphics[width=3.375in]{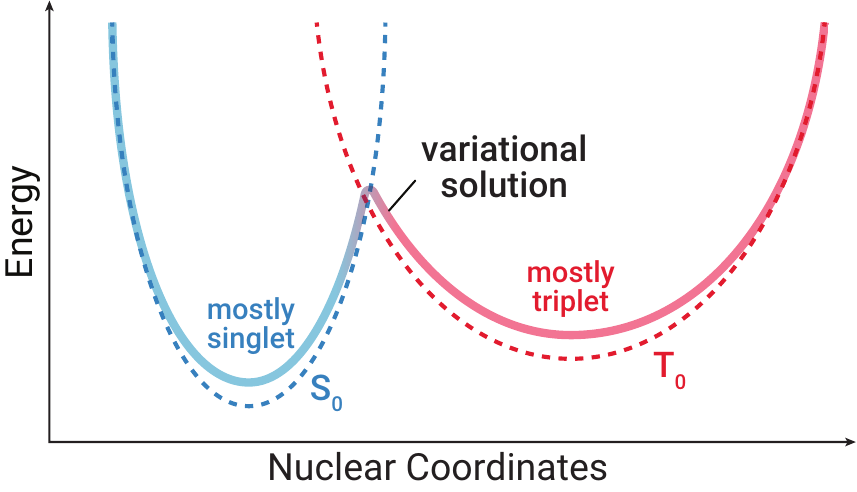}
	\caption{Illustration of crossing between singlet (blue) and triplet (red) eigenvectors (solid lines). A 
		quantum simulation that breaks $S^2$ symmetry and targets the singlet state will produce an unphysical double minimum (dashed line).}
	\label{spin_cross}
\end{figure}

Beyond resource reductions, symmetry-adaptation is also useful for disentangling states with different 
symmetry properties, allowing one to track only those states with desired good quantum numbers.
To demonstrate the usefulness of this aspect, in \cref{spin_cross} we illustrate a case in which the lowest-energy singlet ($\text{S}_0$) and triplet ($\text{T}_0$) potential energy surfaces intersect.
A quantum simulation that does not conserve $S^2$ and targets the $\text{S}_0$ state may erroneously collapse to the $\text{T}_0$ state in the region where the latter is lower in energy, producing a seemingly unphysical potential energy surface.
Symmetry adaptation avoids such issues and allows for the simulation of the lowest-energy state in each symmetry sector, at no additional cost, using standard ground-state quantum algorithms.

Given these practical advantages, we now detail how symmetry-adapted operator pools can be systematically constructed.
The construction of symmetry-adapted operator pools is straightforward when they rely on UCC excitation operators.
The GSD pool is comprised of anti-Hermitian generalized single,
\begin{equation}
	A_p^q = a_p^q - a^p_q \equiv  a^q a_p - a^p a_q,
\end{equation}
and double,
\begin{equation}
	A_{pq}^{rs} = a_{pq}^{rs} - a^{pq}_{rs} \equiv a^r a^s a_q a_p - a^p a^q a_s a_r,
\end{equation}
excitations, where $a_p$ ($a^p$) denotes the annihilation (creation) operator acting on the $p$th spinorbital and all spinorbital indices are generic.
Here and throughout the article, we use upper case $A$ to denote anti-Hermitian fermionic operators.
By creating the same number of electrons as were annihilated, the GSD pool already conserves the total particle number $N$.
The $S_z$ symmetry can be readily enforced by ensuring that all operators in the pool create the same number of $s_z = -\tfrac{1}{2}$ ($\downarrow$) and $s_z = \tfrac{1}{2}$ ($\uparrow$) electrons as were annihilated.
In doing so, there are two distinct types of single excitation operators,
\begin{equation}
	A_{\Pu}^{\Qu} = a^{\Qu}_{\Pu} - a^{\Pu}_{\Qu}
\end{equation}
and
\begin{equation}
	A_{\Pd}^{\Qd} = a^{\Qd}_{\Pd} - a^{\Pd}_{\Qd},
\end{equation}
and three distinct types of double excitation operators,
\begin{equation}
	A_{\Pu\Qu}^{\Ru\Su} = a^{\Ru \Su}_{\Pu\Qu} - a^{\Pu\Qu}_{\Ru\Su},
\end{equation}
\begin{equation}
	A_{\Pd\Qd}^{\Rd\Sd} = a_{\Pd\Qd}^{\Rd\Sd} - a^{\Pd\Qd}_{\Rd\Sd},
\end{equation}
and
\begin{equation}
	A_{\Pu\Qd}^{\Ru\Sd} = a_{\Pu\Qd}^{\Ru\Sd} - a^{\Pu\Qd}_{\Ru\Sd},
\end{equation}
where capital indices designate spatial orbitals, \foreign{i.e.}, we write a spinorbital as $\ket{P\sigma} \equiv \ket{P}\otimes\ket{\sigma}$ with $\sigma\in\{\uparrow,\downarrow\}$.

To impose spatial symmetry in the GSD pool, one retains only those excitation operators that belong to the totally symmetric irreducible representation of the molecular point group.
In the case of the standard Abelian groups commonly used in classical electronic structure methods, namely, $C_i$, $C_2$, $C_s$, $D_2$, $C_{2v}$, $C_{2h}$, and $D_{2h}$, there exists a computationally efficient procedure for identifying the irreducible representation to which an excitation operator belongs.
These groups are examples of elementary Abelian 2-groups, meaning they are Abelian and all elements but the identity have order 2.
As such, they are isomorphic to $\mathbb{Z}_2^n$, namely, $C_i \cong C_2 \cong C_s \cong \mathbb{Z}_2$, $D_2 \cong C_{2v} \cong \mathbb{Z}_2^2$, and $D_{2h} \cong \mathbb{Z}_2^3$, and can be viewed as vector spaces over the binary field $\mathbb{F}_2$.
This isomorphism allows one to map irreducible representations of elementary Abelian 2-groups to ordered $n$-tuples of 0s and 1s, which correspond to binary integers.
Due to the vector space structure, the multiplication of irreducible representations in these groups corresponds to vector addition over $\mathbb{F}_2$, which in turn maps to the bitwise exclusive OR operation.
Therefore, identifying the irreducible representation to which an excitation operator belongs is a highly efficient computational procedure.

To ensure that the GSD pool preserves spin quantum numbers, its elements must commute with $S^2$.
Consequently, the GSD pool needs to be singlet spin-adapted so that its elements cannot change the spin of the state that they are applied to.
For single excitations, we need to couple the spin associated with spatial orbitals $P$ and $Q$ to a singlet.
One readily obtains
\begin{equation}\label{eq:sa_singles}
	A_P^Q = \frac{1}{\sqrt{2}}\left( A_{\Pd}^{\Qd} + A_{\Pu}^{\Qu} \right).
\end{equation}
In the case of doubles, we need to couple the spin associated with spatial orbitals $P$, $Q$, $R$, and $S$ to a final singlet state, which can be accomplished in various ways.
In the adopted coupling scheme \cite{Paldus.1977.10.1002/qua.560110511,Paldus.1977.10.1063/1.434526,Adams.1979.10.1103/PhysRevA.20.1,Chiles.1981.10.1063/1.441643,Paldus.1984.10.1103/PhysRevA.30.2193,Takahashi.1986.10.1063/1.451241,Piecuch.1989.10.1002/qua.560360402,Piecuch.1990.10.1007/BF01119191,Geertsen.1991.10.1016/S0065-3276(08)60364-0,Piecuch.1992.10.1007/BF01113244,Piecuch.1994.10.1063/1.467304,Piecuch.1995.10.1002/qua.560550208,Piecuch.1996.10.1103/PhysRevA.54.1210}, the spin of the spatial orbitals 
with lower, $P$ and $Q$, and upper, $R$ and $S$, excitation indices are coupled independently, yielding intermediate spin quantum numbers $S_{PQ}$ and $S_{RS}$ that can be either 0 or 1.
Since we are interested in a final singlet state, the two intermediate spin quantum numbers must be equal, \foreign{i.e.}, $S_{PQ} = S_{RS} = S_i$.
The singlet spin-adapted double excitation operators can be succinctly expressed as \cite{Paldus.1977.10.1002/qua.560110511,Paldus.1977.10.1063/1.434526,Adams.1979.10.1103/PhysRevA.20.1,Paldus.1984.10.1103/PhysRevA.30.2193,Takahashi.1986.10.1063/1.451241,Piecuch.1989.10.1002/qua.560360402,Piecuch.1990.10.1007/BF01119191,Piecuch.1992.10.1007/BF01113244,Piecuch.1994.10.1063/1.467304,Piecuch.1995.10.1002/qua.560550208,Piecuch.1996.10.1103/PhysRevA.54.1210}
\begin{equation}
	\presuperscript{A}{[S_i]}{_{PQ}^{RS}} = N_{PQ}^{RS} [S_i]^{-\frac{1}{2}} \sum_{\substack{
			\sigma_P, \sigma_Q, \sigma_R, \sigma_S \\
			\in \{-\frac{1}{2}, \frac{1}{2}\}
	}} \sum_{\abs{\sigma}\le S_i} \braket{\tfrac{1}{2}\sigma_P\tfrac{1}{2}\sigma_Q}{S_i \sigma} \braket{\tfrac{1}{2}\sigma_R\tfrac{1}{2}\sigma_S}{S_i \sigma} A_{P\sigma_P Q\sigma_Q}^{R\sigma_R S\sigma_S},
\end{equation}
where $\braket{j_1m_1j_2m_2}{JM}$ denote Clebsch--Gordan coefficients, $[S_i] = 2S_i + 1$ is the spin multiplicity of the common intermediate spin quantum number, and the normalization factor $N_{PQ}^{RS}$ is given by
\begin{equation}
	N_{PQ}^{RS} = \frac{1}{\sqrt{(1+\delta_{PQ})(1+\delta_{RS})}},
\end{equation}
with $\delta_{PQ}$ and $\delta_{RS}$ being Kronecker deltas.
As a result, depending on whether the common intermediate spin state is singlet ($S_i = 0$) or triplet ($S_i = 1$), there are two distinct types of double excitations, given by \cite{Geertsen.1991.10.1016/S0065-3276(08)60364-0}
\begin{equation}\label{eq:sa_doubles_0}
\presuperscript{A}{[0]}{_{PQ}^{RS}} = \frac{1}{2\sqrt{(1+\delta_{PQ})(1+\delta_{RS})}} \left( A_{\Pu \Qd}^{\Ru \Sd} - A_{\Pu \Qd}^{\Rd \Su} - A_{\Pd \Qu}^{\Ru \Sd} + A_{\Pd \Qu}^{\Rd \Su}\right)
\end{equation}
and
\begin{equation}\label{eq:sa_doubles_1}
	\resizebox{\textwidth}{!}{$
	\presuperscript{A}{[1]}{_{PQ}^{RS}} = \dfrac{(1-\delta_{PQ})(1-\delta_{RS})}{\sqrt{3}} \left[ A_{\Pu \Qu}^{\Ru \Su} + A_{\Pd \Qd}^{\Rd \Sd} + \frac{1}{2}\left( A_{\Pu \Qd}^{\Ru \Sd} + A_{\Pu \Qd}^{\Rd \Su} + A_{\Pd \Qu}^{\Ru \Sd} + A_{\Pd \Qu}^{\Rd \Su}\right) \right]
$},
\end{equation}
respectively.
The $(1-\delta_{PQ})(1-\delta_{RS})$ prefactor in \cref{eq:sa_doubles_1} ensures that the upper and lower indices are not repeated.
Note that the operators in \cref{eq:sa_doubles_0,eq:sa_doubles_1} are orthogonally spin-adapted, meaning that, apart from the trivial case in which $\presuperscript{A}{[S_i]}{_{PQ}^{RS}} \ket{\Phi} = 0$, their application to a closed-shell reference Slater determinant $\ket{\Phi}$ leads to a set of orthonormal doubly excited configuration state functions. 

\section{Unitaries of Spin-Adapted Operators}

After constructing the desired symmetry-adapted operator pool, the next step is to design efficient quantum circuits implementing the corresponding unitaries.
Circuit efficiency is typically measured by its depth, \foreign{i.e.}, the number of simultaneously executable layers of gates, and the total count of two-qubit gates, which are the major source of operational noise.
In the case of UCC-based operator pools, the construction of compact circuits enforcing point group, $N$, and $S_z$ symmetry is achieved via the qubit-excitation-based and fermionic-excitation-based formalisms \cite{Yordanov.2020.10.1103/PhysRevA.102.062612,Xia.2021.10.1088/2058-9565/abbc74,Magoulas.2023.10.1021/acs.jctc.2c01016,Sun.2024.2406.11699,Ramoa.2024.2407.08696}.
However, constructing efficient quantum circuits for unitaries of generic spin-adapted operators remains elusive.
This is due to the fact that spin-adapted excitation operators involve linear combinations of spinorbital ones, which generally do not pairwise commute.
As a result, decomposing unitaries generated by spin-adapted operators into finite products of spinorbital unitaries is necessarily approximate.

In the following subsections, we first highlight the limitations of finite-order Trotter--Suzuki decompositions.
We then discuss the mathematical properties of unitaries generated by spin-adapted operators, providing insights into why their exact circuit implementation is challenging on quantum hardware.
Finally, we present exact, closed-form expressions for unitaries generated by singlet spin-adapted double excitations.
%, which can be valuable in guiding the search for efficient quantum circuit implementations.

\subsection{Failure of Finite Trotter--Suzuki Decompositions}

To illustrate the complexity involved with unitaries of spin-adapted operators, we focus on the saGSD pool, which is comprised of totally symmetric, singlet spin-adapted single and double excitations that also enforce $S_z$ symmetry.
Starting with the singles [\cref{eq:sa_singles}], we note that the two constituent terms commute as they operate on spinorbitals with different $S_z$ values, namely, $[A_{\Pu}^{\Qu}, A_{\Pd}^{\Qd}] = 0$.
Consequently, the associated unitary can be exactly expressed as the product of individual spinorbital unitaries,
\begin{equation}
	e^{\frac{\theta}{\sqrt{2}} (A_{\Pd}^{\Qd} + A_{\Pu}^{\Qu})} = e^{\frac{\theta}{\sqrt{2}} A_{\Pd}^{\Qd}} e^{ \frac{\theta}{\sqrt{2}} A_{\Pu}^{\Qu}}.
\end{equation}
As a result, the efficient hardware implementation of unitaries of singlet spin-adapted singles is straightforward.

Before discussing unitaries of singlet spin-adapted doubles, we first partition \cref{eq:sa_doubles_0,eq:sa_doubles_1} based on whether spatial orbital indices $P$ and $Q$, and $R$ and $S$ are unique or repeated.
This yields the following distinct cases:
\begin{equation}\label{eq:sa_doubles_ppqq}	
	A_{PP}^{QQ} = A_{\Pu \Pd}^{\Qu \Qd},
\end{equation}
\begin{equation}\label{eq:sa_doubles_ppqr}
	A_{PP}^{QR} = \tfrac{1}{\sqrt{2}} (A_{\Pu \Pd}^{\Qu \Rd} - 
	A_{\Pu \Pd}^{\Qd \Ru}),
\end{equation}
\begin{equation}\label{eq:sa_doubles_pqrs_0}
	\presuperscript{A}{[0]}{_{PQ}^{RS}} = \tfrac{1}{2}(A_{\Pu \Qd}^{\Ru \Sd} - 
	A_{\Pu \Qd}^{\Rd \Su} - A_{\Pd \Qu}^{\Ru \Sd} + 
	A_{\Pd \Qu}^{\Rd \Su}),
\end{equation}
and
\begin{equation}\label{eq:sa_doubles_pqrs_1}
	\presuperscript{A}{[1]}{_{PQ}^{RS}} = \tfrac{1}{\sqrt{3}} \left[{A_{\Pu \Qu}^{\Ru \Su} +
		A_{\Pd \Qd}^{\Rd \Sd} + \tfrac{1}{2}(A_{\Pu \Qd}^{\Ru 
			\Sd} + A_{\Pu \Qd}^{\Rd \Su} 
		+ A_{\Pd \Qu}^{\Ru \Sd} + A_{\Pd \Qu}^{\Rd \Su})} 
	\right].
\end{equation}
Note that the ``$[0]$'' superscript is dropped in \cref{eq:sa_doubles_ppqq,eq:sa_doubles_ppqr}, as the repeated lower and/or upper excitation indices immediately imply that these singlet spin-adapted double excitations are necessarily of the intermediate singlet type [$S_i = 0$, see \cref{eq:sa_doubles_0}].
Furthermore, perfect pairing double excitations [\cref{eq:sa_doubles_ppqq}] are trivially spin-adapted and, thus, straightforward to implement on quantum hardware.

The double excitations shown in \cref{eq:sa_doubles_ppqr,eq:sa_doubles_pqrs_0,eq:sa_doubles_pqrs_1}, however, consist of linear combinations of noncommuting terms.
Although exact quantum circuit implementations of their associated unitaries are currently unknown, efficient quantum circuits representing approximations can be constructed via finite-order Trotter--Suzuki decompositions.
Depending on whether the unitary is Trotterized before or after the operators are translated from second quantization to the qubit space, a procedure known as fermionic encoding (f.e.) \cite{Jordan.1928.10.1007/BF01331938,Bravyi.2002.10.1006/aphy.2002.6254,Seeley.2012.10.1063/1.4768229,Harrison.2024.2409.04348}, there are two distinct pathways to devise the approximate quantum circuits.
In the first strategy, the spin-adapted unitary is first encoded in qubit space and then Trotterized,
\begin{equation}
	e^{\theta \sum_n c_n A_n} \xrightarrow{\text{f.e.}} e^{\theta \sum_\nu \tilde{c}_\nu P_\nu} \overset{\text{Trot.}}{\approx} \prod_\nu e^{\theta \tilde{c}_\nu P_\nu},
\end{equation}
where $P_\nu$ represents a Pauli string and in the last step we employed a first-order Trotter--Suzuki decomposition for simplicity.
A major issue of this method is that, although Pauli strings originating from a single, anti-Hermitian excitation operator pairwise commute \cite{Romero.2019.10.1088/2058-9565/aad3e4}, strings arising from different operators generally do not.
Consequently, since Pauli strings do not conserve, in general, $N$ and $S_z$, the breaking of these symmetries is possible, depending on the ordering of the exponentials.
This may happen despite the fact that the individual spinorbital excitation operators enforce the $N$ and $S_z$ symmetries.
Note, however, that Pauli strings arising from symmetry-adapted fermionic operators conserve $\mathbb{Z}_2$ symmetries such as point group and fermionic parity, so that the symmetry breaking will only introduce contaminants with $N\pm2$, $N\pm4$,\ldots\ and $S_z\pm 1\,\hbar$, $S_z\pm 2\,\hbar$,\ldots.

In the second approach, the unitary is first Trotterized in the second-quantized space, then translated to the qubit space, and finally factorized,
\begin{equation}
	e^{\theta \sum_n c_n A_n} \overset{\text{Trot.}}{\approx} \prod_n e^{\theta c_n A_n} \xrightarrow{\text{f.e.}} \prod_n e^{\theta \sum_\nu \tilde{c}_\nu^{(n)} P_\nu^{(n)}} = \prod_n \prod_\nu e^{\theta \tilde{c}_\nu^{(n)} P_\nu^{(n)}},
\end{equation}
where, as above, we assumed a first-order Trotter--Suzuki decomposition for simplicity.
Note that the factorization in the last step is exact, as Pauli strings stemming from the same fermionic, anti-Hermitian operator commute.
The advantage of this approach is that by performing the Trotterization in second quantization, the original unitary is approximated by a product of unitaries that individually respect the $N$ and $S_z$ quantum numbers.
To that end, in this work, we adopt the second scheme to maintain these symmetries.

A quick inspection of \cref{eq:sa_doubles_ppqr,eq:sa_doubles_pqrs_0,eq:sa_doubles_pqrs_1} reveals that they involve two [\cref{eq:sa_doubles_ppqr,eq:sa_doubles_pqrs_0}] and three [\cref{eq:sa_doubles_pqrs_1}] groups of noncommuting operators.
Thus, we need to consider product formulas for decompositions of exponentials of two, $e^{\theta (X+Y)}$, and three, $e^{\theta (X+Y+Z)}$, noncommuting terms.
For an exponential of the form $e^{\theta(X+Y)}$ where $X$ and $Y$ do not commute, the first-, second-, and fourth-order Trotter--Suzuki decompositions used in this work are
\begin{equation}\label{eq:trotAB1}
	e^{\theta(X+Y)} \overset{\text{Trot}_1}{\approx} e^{\theta X}e^{\theta Y},
\end{equation}
\begin{equation}\label{eq:trotAB2}
	e^{\theta (X+Y)} \overset{\text{Trot}_2}{\approx} e^{\frac{\theta}{2} X} e^{\theta Y} e^{\frac{\theta}{2} X},
\end{equation}
and
\begin{equation}\label{eq:trotAB4}
	e^{\theta (X+Y)} \overset{\text{Trot}_4}{\approx} e^{\frac{s\theta}{2} X} e^{s\theta Y} e^{\frac{(1-s)\theta}{2} X} e^{(1-2s)\theta Y} e^{\frac{(1-s)\theta}{2} X} e^{s\theta Y} e^{\frac{s\theta}{2} X},
\end{equation}
respectively, where $s = \frac{1}{2-\sqrt[3]{2}}$ \cite{Hatano.2005.10.1007/11526216_2}.
Recall that an $n$th-order Trotter--Suzuki decomposition reproduces the target exponential with leading errors on the order  of $\mathcal{O}(\theta^{n+1})$,
\begin{equation}
	\prod_m e^{c_{2m} \theta X} e^{c_{2m+1} \theta Y} = e^{\theta (X+Y) + \mathcal{O}(\theta^{n+1})}.
\end{equation}
The corresponding expressions in the case of three non-commuting operators are
\begin{equation}
	e^{\theta (X+Y+Z)} \overset{\text{Trot}_1}{\approx} e^{\theta X} e^{\theta Y} e^{\theta Z},
\end{equation}
\begin{equation}
	e^{\theta (X+Y+Z)} \overset{\text{Trot}_2}{\approx} e^{\frac{\theta}{2} X} e^{\frac{\theta}{2} Y} e^{\theta Z} e^{\frac{\theta}{2} Y} e^{\frac{\theta}{2} X},
\end{equation}
and
\begin{equation}
	\resizebox{\textwidth}{!}{$
	e^{\theta (X+Y+Z)} \overset{\text{Trot}_4}{\approx} e^{\frac{s\theta}{2} X} e^{\frac{s\theta}{2} Y} e^{s\theta Z} e^{\frac{s\theta}{2} Y} e^{\frac{(1-s)\theta}{2} X} e^{\frac{(1-2s)\theta}{2} Y} e^{(1-2s)\theta Z} e^{\frac{(1-2s)\theta}{2} Y} e^{\frac{(1-s)\theta}{2} X} e^{\frac{s\theta}{2} Y} e^{s\theta Z} e^{\frac{s\theta}{2} Y} e^{\frac{s\theta}{2} X}
$},
\end{equation}
respectively, with $s$ as before \cite{Barthel.2020.10.1016/j.aop.2020.168165}.
Although more accurate second- and fourth-order Trotterization schemes exist \cite{Barthel.2020.10.1016/j.aop.2020.168165}, they were not considered in this work because they require additional exponential terms, resulting in deeper and longer quantum circuits.

In this study, we gauged the extend to which finite-order Trotter--Suzuki decompositions violate the $S^2$ symmetry.
To illustrate this clearly, we focused on the simplest nontrivial singlet spin-adapted double excitation operator, given by \cref{eq:sa_doubles_ppqr}.
Without loss of generality, we considered the $A_{1\,1}^{3\,5}$ operator from the saGSD operator pool for the H$_6$/STO-6G linear chain.
In our numerical analysis, we constructed the exact matrix exponential of the singlet spin-adapted excitation operator $A_{1\,1}^{3\,5}$ in the Fock space associated with 6 spatial orbitals, labeled 0--5.
In addition, we generated the corresponding matrices from the first-, second-, and fourth-order Trotter--Suzuki approximations of $\exp(\theta A_{1\,1}^{3\,5})$.

To evaluate the quality of these approximate decompositions, we first computed as a function of $\theta$ the Frobenius norm of the difference between the exact exponential and its Trotterized approximations, defined as $\norm{\exp(\theta A_{1\,1}^{3\,5})-\text{Trot}_n}_F$ with $n = 1$, 2, and 4 denoting the first-, second-, and fourth-order Trotter--Suzuki decompositions.
As illustrated in panel (a) of \cref{fig:exact_vs_trot}, as the parameter $\theta$ increases from zero, the difference between the exact exponential and its approximations gradually grows until a maximum error is reached around the $\theta$ values of 4 (first-order) and 5 (second- and fourth-order).
As might have been anticipated, higher-order Trotter--Suzuki decompositions deviate more slowly from the exact exponential as $\theta$ increases.
As shown in the \sm, it is interesting to note that all three error curves exhibit an oscillatory behavior, with the frequency increasing for higher-order Trotter--Suzuki approximations.
Furthermore, with the exception of the trivial case $\theta = 0$, none of the examined product formulas exactly represents $\exp(\theta A_{1\,1}^{3\,5})$ for any values of $\theta$.
Nevertheless, the second- and fourth-order decompositions are faithfully reproducing the target exponential for values of $\theta$ as large as 1.
We, thus, anticipate that these higher-order Trotter--Suzuki formulas are well-suited for instances of weak many-electron correlation effects, characterized by relatively small cluster amplitudes.
\begin{figure}[!h]
	\centering
	\includegraphics[width=3.375in]{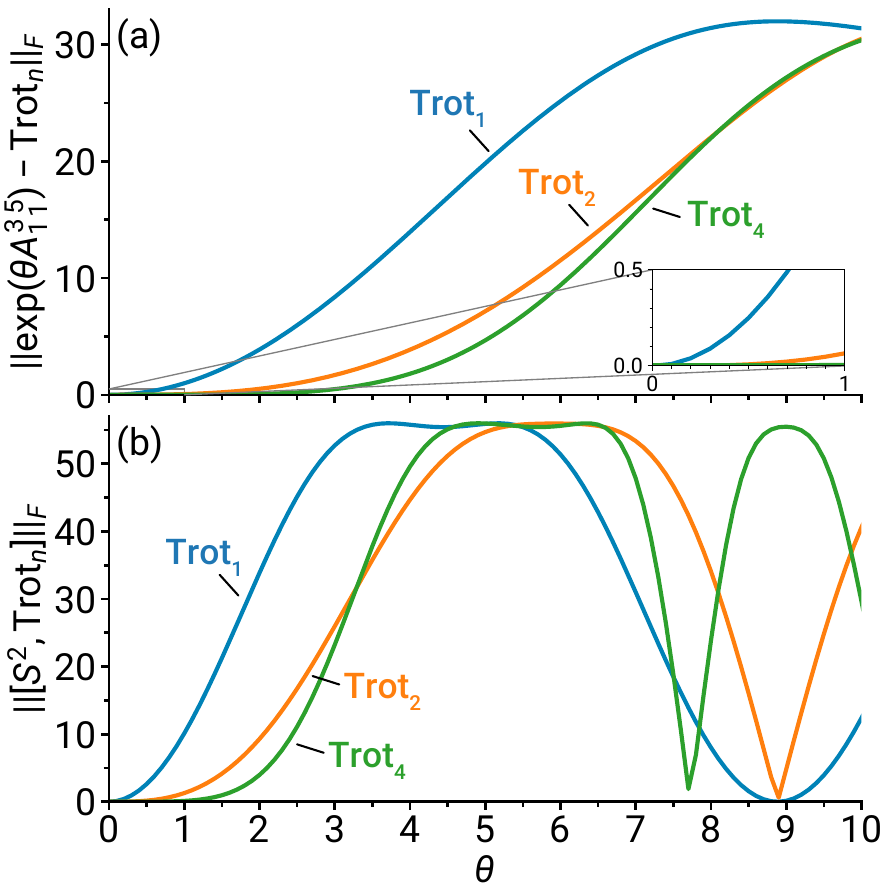}
	\caption{
		(a) The Frobenius norm of the difference between the exact unitary $\exp({\theta A_{1\,1}^{3\,5}})$ and its first-, second-, and fourth-order Trotter--Suzuki approximations as a function of $\theta$.
		(b) The Frobenius norm of the commutator between the total spin squared operator ($S^2$) and the first-, second-, and fourth-order Trotter--Suzuki decompositions of the unitary $\exp({\theta A_{1\,1}^{3\,5}})$ as a function of $\theta$. Here, $A_{1\,1}^{3\,5}$ is the anti-Hermitian, singlet spin-adapted double excitation operator involving spatial orbitals $(1,1)\leftrightarrow(3,5)$ (orbital indices start at 0). All operator matrices were constructed in the Fock space of six spatial orbitals.
	}
	\label{fig:exact_vs_trot}
\end{figure}

Despite the inability of finite-order Trotter--Suzuki decompositions to accurately represent the target unitary, it is still worth examining the degree to which $S^2$ symmetry is violated.
To that end, we constructed the matrix representation of the $S^2$ operator in the same Fock space and computed the Frobenius norm of the commutator $[S^2, \text{Trot}_n]$.
As shown in panel (b) of \cref{fig:exact_vs_trot}, the observed behavior for $\theta \in [0,4]$ closely mirrors the pattern seen previously in panel (a).
The major difference is that all three curves wildly oscillate for larger values of $\theta$.
Moreover, as shown in the \sm, the first- and second-order Trotter--Suzuki approximations to $\exp(\theta A_{1\,1}^{3\,5})$ appear to be spin-adapted for $\theta = k2\sqrt{2}\pi$, $k\in\mathbb{Z}$.
However, by examining \cref{eq:trotAB1,eq:trotAB2}, we realize that for these values of $\theta$ both of these approximations reduce to the identity operator and, thus, trivially commute with $S^2$.
As might have been anticipated, the spin symmetry breaking introduced by the higher-order Trotter--Suzuki decompositions appears to be negligible for $\theta<1$.

%To further quantify the degree of $S^2$-symmetry breaking induced by Trotterization, we performed the following numerical test.
%First, we constructed a state $\ket*{D}$ that is an equal superposition of all singlet spin-adapted configuration state functions with 6 electrons (half-filling).
%Subsequently, we acted on this state with the three product formulas approximating $\exp(\theta A_{1\,1}^{3\,5})$, $\ket*{\tilde{D}_n} = \text{Trot}_n \ket*{D}$.
%Finally, we computed the expectation value and standard deviation of $S^2$ with respect to the states $\ket*{\tilde{D}_n}$ as functions of the amplitude $\theta$.
%As shown in panel (c) of \cref{fig:exact_vs_trot}...

Further insights into the challenges faced by finite-order Trotter--Suzuki decompositions in preserving the $S^2$ symmetry can be gained by expressing them in a closed form.
Unitaries generated by an anti-Hermitian linear combination of fermionic strings can be exactly expressed as \cite{Evangelista.2019.10.1063/1.5133059,Filip.2020.10.1063/5.0026141,Chen.2021.10.1021/acs.jctc.0c01052,Rubin.2021.10.22331/q-2021-10-27-568,Xu.2023.10.3390/sym15071429}
\begin{equation}\label{eq:exp_closed}
	e^{\theta A} = I + \sin(\theta)A + [1-\cos(\theta)]A^2,
\end{equation}
where $I$ is the identity, $A = F-F^\dagger$ is the anti-Hermitian generator, and $F$ is a product of second-quantized annihilation, creation, and number operators.
Note that $F$ cannot be a product of number operators only since in that case $F-F^\dagger = 0$.
Using \cref{eq:exp_closed} and after a few algebraic manipulations, the first-order Trotter--Suzuki decomposition of $\exp(\theta A_{PP}^{QR}/\sqrt{2})$ is expressed in the closed-form
\begin{equation}\label{eq:trot1_closed}
	\begin{split}
		&e^{\frac{\theta}{\sqrt{2}} A_{\Pu\Pd}^{\Qu\Rd}} e^{-\frac{\theta}{\sqrt{2}} A_{\Pu\Pd}^{\Qd\Ru}} ={}
		\textcolor{myblue}{I}\\
		&+ \sin{\left(\frac{\theta}{\sqrt{2}} \right)} \left(\textcolor{myblue}{A_{\Pu\Pd}^{\Qu\Rd} - A_{\Pu\Pd}^{\Qd\Ru}}\right)\\
		&+\frac{1}{2}\sin\left(\frac{\theta}{\sqrt{2}}\right)\left[\cos\left(\frac{\theta}{\sqrt{2}}\right) -1\right]\textcolor{myblue}{\left(A_{\Pu\Pd}^{\Qu\Rd} - A_{\Pu\Pd}^{\Qd\Ru} \right)\left(\bar{n}_{\Qd \Ru} + n_{\Qd \Ru}\right) \left(\bar{n}_{\Qu \Rd} + n_{\Qu \Rd}\right)}\\
		&+ \left[\cos{\left(\frac{\theta}{\sqrt{2}} \right)} - 1\right]^2 \textcolor{myblue}{\left(\bar{n}_{P\uparrow P\downarrow} n_{Q\uparrow Q\downarrow R\uparrow R\downarrow} + \bar{n}_{Q\uparrow Q\downarrow R\uparrow R\downarrow} n_{P\uparrow P\downarrow}\right)}\\
		&+ \sin^{2}{\left(\frac{\theta}{\sqrt{2}} \right)} \textcolor{myorange}{\left(a^{Q\downarrow R\uparrow}_{Q\uparrow R\downarrow} n_{P\uparrow P\downarrow} + a^{Q\uparrow R\downarrow}_{Q\downarrow R\uparrow} \bar{n}_{P\uparrow P\downarrow}\right)}\\
		&+\left[\cos{\left(\frac{\theta}{\sqrt{2}} \right)} - 1\right] \textcolor{myorange}{\left[\bar{n}_{P\uparrow P\downarrow}\left( n_{Q\downarrow R\uparrow} + n_{Q\uparrow R\downarrow}\right) + \left(\bar{n}_{Q\downarrow R\uparrow} + \bar{n}_{Q\uparrow R\downarrow}\right) n_{P\uparrow P\downarrow}\right]}\\
		&+\frac{1}{2}\sin\left(\frac{\theta}{\sqrt{2}}\right)\left[\cos\left(\frac{\theta}{\sqrt{2}}\right) -1\right]\textcolor{myorange}{\left( A_{\Pu\Pd}^{\Qu\Rd} + A_{\Pu\Pd}^{\Qd\Ru} \right)\left(\bar{n}_{\Qd \Ru} - n_{\Qd \Ru}\right) \left(\bar{n}_{\Qu \Rd} - n_{\Qu \Rd}\right)}
	\end{split}
\end{equation}
with $n_{p_1 \ldots p_i} \equiv n_{p_1}\cdots n_{p_i}$ and $\bar{n}_{p_1 \ldots p_i} \equiv \bar{n}_{p_1}\cdots \bar{n}_{p_i}$ denoting collections of particle and hole number operators, respectively.
The failure of the first-order product formula to preserve spin symmetry is striking.
In addition to the singlet spin-adapted contributions, marked in blue, \cref{eq:trot1_closed} contains terms that are not singlet spin-adapted, denoted in orange.
For example, in the last term in \cref{eq:trot1_closed}, we see the appearance of $A_{\Pu\Pd}^{\Qu\Rd} + A_{\Pu\Pd}^{\Qd\Qu}$ that is triplet spin-adapted.
In the \sm, we give an equivalent representation in terms of singlet and triplet spin-adapted operators and their products.
In doing so, it becomes apparent that up to quintet spin-adapted operators contribute to this expression.
For the first-order Trotter--Suzuki decomposition of $\exp(\theta A_{PP}^{QR}/\sqrt{2})$ to be singlet spin-adapted, the offending terms must vanish.
A quick inspection of \cref{eq:trot1_closed} immediately reveals that this occurs for $\theta = k2\sqrt{2}\pi$, $k\in\mathbb{Z}$.
This analytical result is in complete agreement with our numerical observations, as shown in panel (b) of \cref{fig:exact_vs_trot} and in the \sm.
However, as already noted above, for these values of $\theta$, the first-order product formula reduces to the identity operator, and, thus, it is trivially singlet spin-adapted.
As shown in the \sm, the closed-form expressions for the second- and fourth-order Trotter--Suzuki decompositions of $\exp(\theta A_{PP}^{QR}/\sqrt{2})$ involve additional terms, some of which are singlet spin-adapted and some that are not.
Furthermore, the trigonometric functions multiplying the various operators become increasingly more complex.

\subsection{Periodicity}

The mathematical properties of exponentials of 
spin-adapted operators can shed additional light into why it is challenging to find efficient circuit representations and good approximations based on product formulas. 
Based on \cref{eq:exp_closed}, unitaries generated by anti-Hermitian combinations of spinorbital operators are periodic functions of their corresponding arguments, with period $2\pi$.
The question then arises whether unitaries of spin-adapted operators are also periodic and, if so, what is the corresponding period.

As before, we focus on the simplest non-trivial singlet spin-adapted double excitation operator, namely, $A_{PP}^{QR}$.
Let us assume that $\exp(\theta A_{PP}^{QR})$ is a periodic function of $\theta$ with period $T$.
Using the fact that $f(\theta + T) = f(\theta)$, one can readily obtain that $e^{T A_{PP}^{QR}} = I$.
As a result, if $\exp(\theta A_{PP}^{QR})$ is periodic, it must become the identity operator at regular intervals corresponding to the period.

\begin{figure}[!h]
\centering
	\includegraphics[width=3.375in]{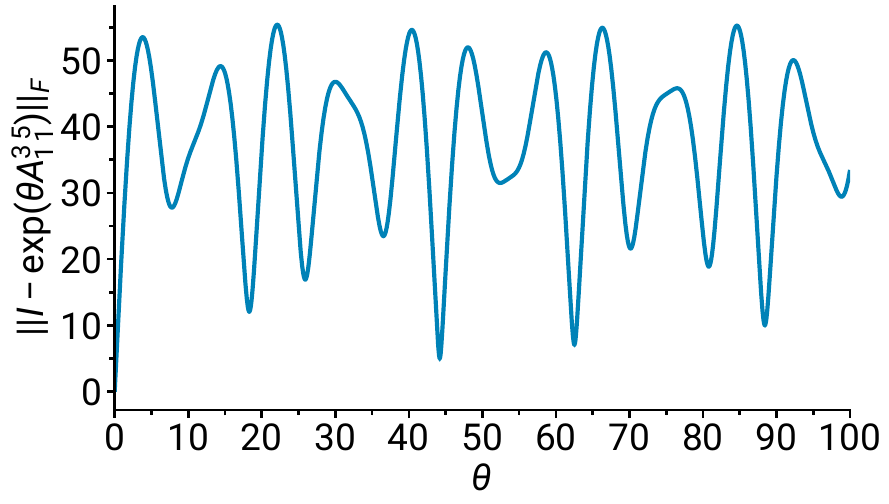}
	\caption{The Frobenius norm of the difference between the identity matrix $I$ and the exact unitary $\exp({\theta A_{1\,1}^{3\,5}})$ as a function of $\theta$. Here, $A_{1\,1}^{3\,5}$ is the anti-Hermitian, singlet spin-adapted double excitation operator involving spatial orbitals $(1,1)\leftrightarrow(3,5)$ (orbital indices start at 0). All operator matrices were constructed in the Fock space of six spatial orbitals.}
	\label{fig:periodicity}
\end{figure}
To numerically verify the periodicity of $\exp(\theta A_{PP}^{QR})$, we worked as follows.
Without loss of generality, we constructed the matrix representation of the singlet spin-adapted excitation operator $A_{1\,1}^{3\,5}$ in the Fock space generated by 6 spatial orbitals, labeled 0--5.
Subsequently, for a wide range of $\theta$ values, we generated the exact matrix exponential $\exp(\theta A_{1\,1}^{3\,5})$ and computed the Frobenius norm $\norm{I - \exp(\theta A_{1\,1}^{3\,5})}_F$, where, in this context, $I$ denotes the identity matrix.
If the matrix exponential is periodic, then the norm should become 0 at regular intervals corresponding to the period $T$.
As shown in \cref{fig:periodicity}, the spin-adapted unitary does not become the entity in the large interval $\theta\in[0,100]$.
Although the unitary is not periodic, it appears to be almost periodic.
Recall that, given $\epsilon>0$, a function $f(\theta)$ is almost periodic with $\epsilon$-period $T$ if and only if $\norm{f(\theta + T) - f(t)}\le\epsilon$ \cite{Amerio.1971.10.1007/978-1-4757-1254-4}.

Despite its usefulness, the above numerical test is not conclusive, as there is always the possibility that the period exceeds the examined region of $\theta$ values.
As shown in the \sm, a unitary is periodic if and only if the ratio of the nonzero eigenvalues of its generator is rational.
Since the examined spin-adapted unitaries are anti-Hermitian, their eigenvalues are purely imaginary and come in pairs that differ in their sign.
In our numerical example, the non-zero eigenvalues of the $A_{1\,1}^{3\,5}$ matrix are $\pm i$ and $\pm 1/\sqrt{2}$.
This analytic approach also proves that indeed the matrix exponential $\exp(\theta A_{1\,1}^{3\,5})$ is not periodic.
A proof that does not rely on the representation of unitaries generated by singlet spin-adapted operators in a finite basis is provided in the next section, where we examine their exact, closed-form expressions.

The lack of periodicity and the possibility that unitaries of spin-adapted operators are almost periodic functions add an extra layer of complexity when seeking product decompositions.
Indeed, as can be seen in \cref{eq:trotAB1,eq:trotAB2} and in the \sm, the first- and second-order Trotter--Suzuki decompositions of $\exp(\theta A_{PP}^{QR})$ are periodic functions with periods $2\sqrt{2}\pi$ and $4\sqrt{2}\pi$, respectively.
As such, they are not suitable approximations to the spin-adapted unitary.
Moving on to the fourth-order approximation given in \cref{eq:trotAB4}, we observe that it involves a product of periodic functions with incommensurate periods and, thus, is an almost-periodic function.
Although this is a step in the right direction, even the fourth-order approximant is not flexible enough to reproduce the complex behavior of the full singlet spin-adapted unitary (see \cref{fig:exact_vs_trot}).

At this point it is worth mentioning that similar problems are anticipated to arise in approximations based on other types of product decompositions.
For example, the Zassenhaus formula is the dual to the Baker--Campbell--Hausdorff identity and reads
\begin{equation}
	e^{\theta(X+Y)} = e^{\theta X} e^{\theta Y} e^{-\frac{\theta^2}{2} [X,Y]} e^{\frac{\theta^3}{3!}([[Y,[Y,X,Y]]] + [X,[X,Y]])}\cdots,
\end{equation}
(see \cite{Taube.2006.10.1002/qua.21198} for an application to UCC theory).
In addition to breaking spin symmetry, an additional complexity arises when a spin-adapted unitary is approximated by a finite-order Zassenhaus decomposition.
For exponentials having a linear combination of commutators in the exponent, it may not be possible to construct the associated efficient quantum circuits, similar to the case of the original spin-adapted unitary.

\subsection{Closed-Form Expressions}

In the previous sections, we highlighted the difficulties with approximating unitaries generated by singlet spin-adapted operators as finite products of unitaries of spinorbital operators.
In this section we take a different route.
Encouraged by the fact that exponentials of spinorbital operators can be expressed in closed form [\cref{eq:exp_closed}],
we seek closed-form expressions of unitaries whose generators are given by \cref{eq:sa_doubles_ppqr,eq:sa_doubles_pqrs_0,eq:sa_doubles_pqrs_1}.

Upon Taylor expansion, one is confronted with the various powers of the pertinent generator.
However, the powers of these spin-adapted operators generate a finite algebra, meaning that, for high enough powers, the same set of spinorbital operators will be appearing.
This directly implies that a singlet spin-adapted unitary can be decomposed as a linear combination of a finite number of fermionic operators.
Each of these operators will be multiplied by an infinite power series in the parameter $\theta$.
The question then arises whether these series can be expressed in a closed form.

By developing new tools for the symbolic manipulation of fermionic operators based on Sympy \cite{Meurer.2017.10.7717/peerj-cs.103}, we generated the finite algebra resulting from the various powers of the given generators.
Subsequently, by interfacing our code with Mathematica \cite{Mathematica}, we found the closed-form expressions for the $\theta$-dependent coefficients.
The closed-form expression in terms of spinorbital operators for the simplest, non-trivial spin-adapted unitary considered in this work is given by
\begin{equation}\label{eq:sa_unitary_closed1}
	\begin{split}
		&e^{\theta A_{PP}^{QR}} ={} I\\
		&+\sin\left(\frac{\theta}{\sqrt{2}}\right) \left( A_{\Pu\Pd}^{\Qu\Rd} - A_{\Pu\Pd}^{\Qd\Ru} \right)\\
		&+\left[ \cos(\frac{\theta}{\sqrt{2}}) -1 \right] \left[\left( \bar{n}_{\Qu \Rd} + \bar{n}_{\Qd \Ru} - a_{\Qu \Rd}^{\Qd \Ru} - a_{\Qd \Ru}^{\Qu \Rd} \right) n_{\Pu \Pd}\right.\\
		&\phantom{+\left[ \cos(\frac{\theta}{\sqrt{2}}) -1 \right]}\!\left.+\left( n_{\Qu \Rd} + n_{\Qd \Ru} -a_{\Qu \Rd}^{\Qd \Ru} - a_{\Qd \Ru}^{\Qu \Rd} \right) \bar{n}_{\Pu \Pd} \right] \\
		&+\left[ \frac{1}{\sqrt{2}}\sin\left( \theta \right) - \sin\left( \frac{\theta}{\sqrt{2}}\right) \right] \left( A_{\Pu\Pd}^{\Qu\Rd} - A_{\Pu\Pd}^{\Qd\Ru} \right) \left(\bar{n}_{\Qd \Ru} + n_{\Qd \Ru}\right) \left(\bar{n}_{\Qu \Rd} + n_{\Qu \Rd}\right) \\
		&+\left[\cos(\theta) - 2\cos(\frac{\theta}{\sqrt{2}}) + 1 \right] \left[ \bar{n}_{\Pu \Pd} n_{\Qu \Qd \Ru \Rd} + \bar{n}_{\Qu \Qd \Ru \Rd} n_{\Pu \Pd} \vphantom{\frac{1}{2}}\right. \\
		&\left.+ \frac{1}{2} \left( \bar{n}_{\Pu \Pd} + n_{\Pu \Pd}\right)\left( \bar{n}_{\Qu \Rd}n_{\Qd \Ru} + \bar{n}_{\Qd \Ru}n_{\Qu \Rd} -a_{\Qu \Rd}^{\Qd \Ru} - a_{\Qd \Ru}^{\Qu \Rd} \right)\right],
	\end{split}
\end{equation}
while the closed-form expressions for the unitaries generated by the more complex spin-adapted double excitations can be found in the \sm.
As might have been anticipated, the various terms in \cref{eq:sa_unitary_closed1} can be partitioned into anti-Hermitian and Hermitian ones, arising respectively from even and odd powers of the anti-Hermitian generator [\cref{eq:sa_doubles_ppqr}].
The anti-Hermitian terms contain the spin-adapted generator multiplied by, at most, a linear combination of number operators.
The Hermitian terms involve number operators and the Hermitian operator that flips the spins of spatial orbitals $Q$ and $R$.
Furthermore, the unitary is exactly represented as a linear combination of trigonometric functions with incommensurate periods.
As such, the spin-adapted unitary is an almost-periodic function of the parameter $\theta$ \cite{Amerio.1971.10.1007/978-1-4757-1254-4}, in complete agreement with the results of our numerical analysis in the previous section.

The existence of closed-form expressions for spin-adapted unitaries, such as the one shown in \cref{eq:sa_unitary_closed1}, provides yet another avenue to gauge the utility of finite Trotter--Suzuki decompositions.
Indeed, even the first-order approximation of \cref{eq:trot1_closed} is able to capture a few of the operators appearing in \cref{eq:sa_unitary_closed1}, albeit with potentially different coefficients.
However, as already mentioned above, product approximations result in additional terms that are not singlet spin adapted.
As the order of the approximation is increased, product formulas will result in additional terms until the algebra is saturated.
After this point, further increasing the order of the approximation will result in modifications to the trigonometric functions multiplying the various terms.

Even though the simplest, nontrivial singlet spin-adapted generator of \cref{eq:sa_doubles_ppqr} is far too complex for a straightforward quantum gate decomposition, the closed-form expression shown in \cref{eq:sa_unitary_closed1} enables its hardware implementation via a different route.
The right-hand side of \cref{eq:sa_unitary_closed1} can be expressed as a linear combination of Pauli strings, after translating the second-quantized operators to the qubit space.
Consequently, the spin-adapted unitary can be exactly implemented on a quantum device as a linear combination of unitaries (LCU) \cite{Childs.2012.1202.5822,Berry.2015.10.1103/PhysRevLett.114.090502}.
The efficiency of the implementation can be improved by taking advantage of block encoding \cite{Gilyen.2019.10.1145/3313276.3316366} and qubitization \cite{Low.2017.10.1103/PhysRevLett.118.010501,Low.2019.10.22331/q-2019-07-12-163} algorithms.
In the former, the LCU is embedded as a sub-block of a larger unitary matrix, while in the latter, qubitization transforms the LCU into a quantum walk operator whose controlled phase rotations efficiently encode the eigenstructure of the target unitary, enabling its precise implementation with optimal scaling in gate count and error bounds.

Despite the usefulness of \cref{eq:sa_unitary_closed1}, it is not immediately obvious that the various terms appearing in its right-hand side are singlet spin-adapted.
After a few algebraic manipulations, \cref{eq:sa_unitary_closed1} can be written in terms of singlet spin-adapted operators as
\begin{equation}\label{eq:sa_unitary_closed2}
	\resizebox{\textwidth}{!}{$
	\begin{split}
		&e^{\theta A_{PP}^{QR}} ={} I\\
		&+\sqrt{2}\sin\left(\frac{\theta}{\sqrt{2}}\right) A_{PP}^{QR} \\
		&+\left[ \cos(\frac{\theta}{\sqrt{2}}) -1 \right] \left[\left( 2-\sqrt{2}a_Q^Q-\sqrt{2}a_R^R+2\presuperscript{a}{[0]\,}{^{QR} _{QR}} \right) n_{PP} +2\presuperscript{a}{[0]\,}{^{QR} _{QR}} \bar{n}_{PP} \right] \\
		&+\left[ \sin\left( \theta \right) - \sqrt{2}\sin\left( \frac{\theta}{\sqrt{2}}\right) \right] A_{PP}^{QR} \left[\left( 2-\sqrt{2}a_Q^Q-\sqrt{2}a_R^R+2\presuperscript{a}{[0]\,}{^{QR} _{QR}} \right) n_{PP} +2\presuperscript{a}{[0]\,}{^{QR} _{QR}} \bar{n}_{PP} -1\right] \\
		&+\left[\cos(\theta) - 2\cos(\frac{\theta}{\sqrt{2}}) + 1 \right] \left[ \bar{n}_{PP} n_{QQRR} + \bar{n}_{QQRR} n_{PP} \vphantom{\frac{1}{2}}\right. \\
		&\left.+ \frac{1}{2} \left( \bar{n}_{PP} + n_{PP}\right)\left( 2n_{QQRR} - \sqrt{2}n_{QQ}a_R^R - \sqrt{2}n_{RR}a_Q^Q + 2 \presuperscript{a}{[0]\,}{^{QR} _{QR}}\right)\right].
	\end{split}
$}
\end{equation}
In addition to the $A_{PP}^{QR}$ generator, \cref{eq:sa_unitary_closed2} contains ``perfect-pairing'' number operators, such as $n_{PP} = n_{\Pu\Pd} = a_{\Pu\Pd}^{\Pu\Pd}$, the singlet spin-adapted single excitations $a_Q^Q$ and $a_R^R$, defined as
\begin{equation}
	a_Q^Q =\frac{1}{\sqrt{2}}\left(n_{\Qd} + n_{\Qu} \right)
\end{equation}
and similar for $a_R^R$, and the singlet spin-adapted double excitation going through an intermediate singlet involving the spatial orbitals $Q$ and $R$, given by
\begin{equation}
	\presuperscript{a}{[0]\,}{^{QR} _{QR}} = n_{\Qu\Rd} - a_{\Qu\Rd}^{\Qd\Ru} - a_{\Qd\Ru}^{\Qu\Rd} + n_{\Qd\Ru}.
\end{equation}

Arguably, the most compact closed form of the spin-adapted unitary is obtained by expressing it in terms of the various powers of $A_{PP}^{QR}$.
In doing so, we arrive at
\begin{equation}\label{eq:sa_unitary_closed3}
	\begin{split}
		e^{\theta A_{PP}^{QR}} ={}& I\\
		&+\left[2\sqrt{2}\sin\left(\frac{\theta}{\sqrt{2}}\right) - \sin\left(\theta\right)\right]A_{PP}^{QR}\\
		&+\left[\cos\left(\theta\right) -4\cos(\frac{\theta}{\sqrt{2}}) +3 \right]{A_{PP}^{QR}}^2\\
		&-2\left[\sin\left( \theta \right) - \sqrt{2} \sin\left( \frac{\theta}{\sqrt{2}}\right)\right]{A_{PP}^{QR}}^3 \\
		&+2\left[\cos(\theta) - 2\cos(\frac{\theta}{\sqrt{2}}) + 1 \right] {A_{PP}^{QR}}^4,
	\end{split}
\end{equation}
which is the complete analog to the well-known formula for spinorbital operators, shown in \cref{eq:exp_closed}.
The form of \cref{eq:sa_unitary_closed3} can be rationalized by considering the powers of $A_{PP}^{QR}$ and the algebras that they generate.
As it turns out, the third and fourth powers of $A_{PP}^{QR}$ already span the algebra associated with the anti-Hermitian and Hermitian components of the corresponding unitary, respectively.
Higher powers simply result in the same sets of operators, albeit with different multiplicative coefficients.

\section{Summary}

In this work, we highlighted substantial shortcomings of finite product formulas in accurately approximating spin-adapted unitaries and retaining good spin quantum numbers.
We demonstrated numerically that even high-order Trotter--Suzuki decompositions, up to fourth order, fail to capture the complex behavior of unitaries generated by singlet spin-adapted double excitations as a function of their associated parameters.
In addition, we showed numerically that all examined product formulas break spin symmetry.
We explained this behavior analytically by proving that product formulas inherently introduce triplet and quintet spin-adapted terms, thus leading to a loss of $S^2$ symmetry.

In comparing unitaries generated by spinorbital operators to those generated by spin-adapted ones, we found that the former are periodic functions of their parameters, while the latter are not.
Both numerical and analytical evidence, based on matrix representations, demonstrated that the functional dependence of spin-adapted unitaries on their parameters is consistent with almost periodic functions.
This finding further underscores the challenges associated with devising finite product formulas capable of closely reproducing such complex behaviors.

We also derived closed-form expressions for unitaries generated by singlet spin-adapted double excitation operators.
These expressions are linear combinations of trigonometric functions with incommensurate periods, providing a conclusive analytical proof that spin-adapted unitaries are almost periodic functions.
The closed-form expressions derived in this work can facilitate the efficient implementation of spin-adapted unitaries on quantum hardware.
One promising route is to implement these unitaries as a linear combinations of simpler unitaries, enabling an embarrassingly parallelizable approach.
Alternatively, these expressions serve as a valuable guide for designing novel approximation strategies based on improved product formulas.
Finally, the availability of closed-form expressions may substantially advance the efforts toward discovering efficient, exact quantum circuit representations of spin-adapted unitaries.

\section*{Acknowledgements}

We dedicate this paper to Professor Piotr Piecuch on the occasion of his 65th birthday. The authors thank Drs.\ Karol Kowalski and Nicholas Bauman for the invitation to contribute to the Special Issue of Molecular Physics in honour of Professor Piotr Piecuch.

\section*{Funding}

This work was supported by the U.S.\ Department of Energy under Award No.\ DE-SC0019374.

\clearpage

\appendix

\section*{Supplementary Material}

This {\sm} document is organized as follows.
In \cref{sec:numerical}, we provide, in graphical form, the results of additional numerical simulations highlighting the failures of first-, second-, and fourth-order Trotter-Suzuki decompositions in representing the singlet spin-adapted unitaries.
\cref{sec:trotn_closed} contains the closed-form expressions for the first- and second-order Trotterization schemes of $\exp(\theta A_{PP}^{QR})$.
In \cref{sec:periodicity}, we provide the proof that a unitary is periodic if and only if the ratio of the nonzero eigenvalues of its generator are rational.
\cref{sec:sa_doubles_closed} contains the closed-form expressions for unitaries generated by singlet spin-adapted double excitations going through intermediate singlets and triplets.

\section{Results of Additional Numerical Simulations}\label{sec:numerical}
\begin{figure}[!h]
	\centering
	\includegraphics[width=\textwidth]{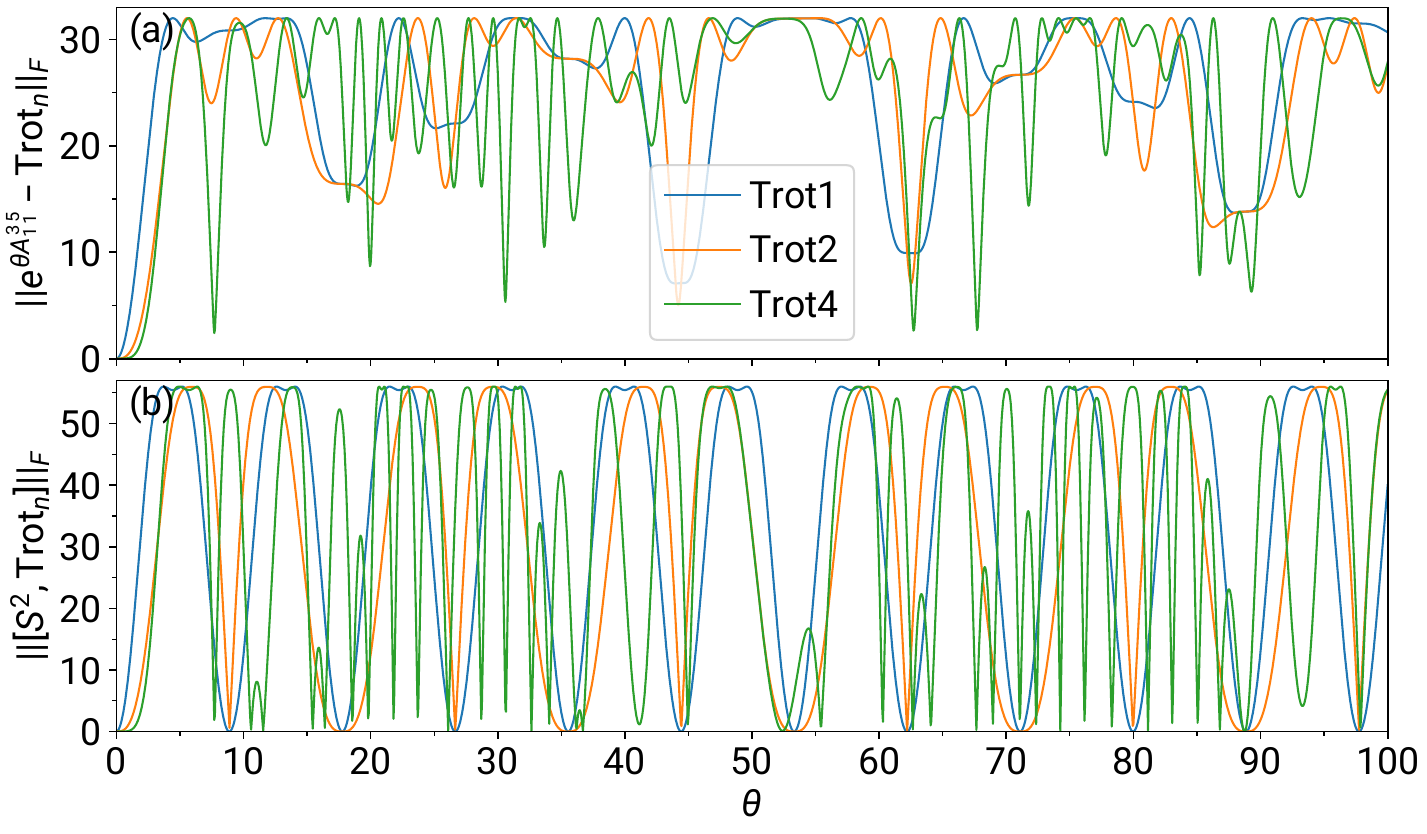}
	\caption{
		(a) The Frobenius norm of the difference between the exact unitary $\exp({\theta A_{1\,1}^{3\,5}})$ and its first-, second-, and fourth-order Trotter-Suzuki approximations as a function of $\theta$.
		(b) The Frobenius norm of the commutator between the total spin squared operator ($S^2$) and the first-, second-, and fourth-order Trotter-Suzuki decompositions of the unitary $\exp({\theta A_{1\,1}^{3\,5}})$ as a function of $\theta$. Here, $A_{1\,1}^{3\,5}$ is the anti-Hermitian, singlet spin-adapted double excitation operator involving spatial orbitals $(1,1)\leftrightarrow(3,5)$ (orbital indices start at 0). All operator matrices were constructed in the Fock space of six spatial orbitals.
	}
	\label{fig:exact_vs_trot}
\end{figure}

\section{Closed-Form Expressions for the First- and Second-Order Decompositions of $\boldsymbol{\exp(\theta A_{PP}^{QR})}$}\label{sec:trotn_closed}

In \cref{eq:trot1_closed} of the main text, we provided a closed-form expression for the first-order Trotter-Suzuki decomposition of the singlet spin-adapted unitary $\exp(\theta A_{PP}^{QR})$ in terms of spinorbital operators.
Here, we provide an equivalent representation in terms of $S\equiv A_{PP}^{QR}$ and its triplet spin-adapted counterpart, namely,
\begin{equation}
	T = \frac{1}{\sqrt{2}}\left( A_{\Pu\Pd}^{\Qu\Rd} + A_{\Pu\Pd}^{\Qd\Ru} \right).
\end{equation}
We begin by noting that the associated spinorbital excitation operators can be expressed in terms of the spin-adapted ones as
\begin{equation}
	A_{\Pu\Pd}^{\Qu\Rd} = \frac{1}{\sqrt{2}}\left( S + T\right)
\end{equation}
and
\begin{equation}
	A_{\Pu\Pd}^{\Qd\Ru} = -\frac{1}{\sqrt{2}}\left(S - T\right).
\end{equation}
Using the above relations in combination with \cref{eq:exp_closed} of the main text, we obtain
\begin{equation}
	\begin{split}
		e^{\frac{\theta}{\sqrt{2}} A_{\Pu\Pd}^{\Qu\Rd}} e^{-\frac{\theta}{\sqrt{2}} A_{\Pu\Pd}^{\Qd\Ru}} ={}& I\\
		&+ \sqrt{2}\sin\left(\frac{\theta}{\sqrt{2}}\right)S\\
		&+\left\{\sqrt{2}\left[ 1 - \cos\left( \frac{\theta}{\sqrt{2}}\right)\right]+\frac{1}{2}\sin^2\left( \frac{\theta}{\sqrt{2}}\right)\right\} S^2\\
		&+\sin\left(\frac{\theta}{\sqrt{2}}\right)\left[1 - \cos\left( \frac{\theta}{\sqrt{2}} \right) \right]S^3\\
		&+\frac{1}{2}\left[1 - \cos\left( \frac{\theta}{\sqrt{2}} \right) \right]^2 S^4\\
		&+\left\{\sqrt{2}\left[ 1 - \cos\left( \frac{\theta}{\sqrt{2}}\right)\right]-\frac{1}{2}\sin^2\left( \frac{\theta}{\sqrt{2}}\right)\right\} T^2\\
		&+\sqrt{2}\left[ 1 - \cos\left( \frac{\theta}{\sqrt{2}}\right)\right] \left(TS-ST \right)\\
		&+\sin\left(\frac{\theta}{\sqrt{2}}\right)\left[1 - \cos\left( \frac{\theta}{\sqrt{2}} \right) \right] \left( TS^2 - TST - S^2 T \right)\\
		&+\frac{1}{2}\left[1 - \cos\left( \frac{\theta}{\sqrt{2}} \right) \right]^2 \left( S T S^2 - S T^2 S + S T^3 - S^2 T S - S^3 T + T S T^2\right.\\
		&\phantom{+\frac{\left[1 - \cos\left( \frac{\theta}{\sqrt{2}} \right) \right]^2}{2} +{}}\left.- T S^2 T + T S^3 - T^2 S T - T^3 S + T^4  \right)
	\end{split}
\end{equation}
The above expression contains two types of terms.
The first one involves only $S$ and its powers and, thus, it is singlet spin-adapted.
The second family of terms contains products of $S$ and $T$ and their powers.
It can be shown that powers of the anti-Hermitian triplet spin-adapted operator $T$ can be decomposed into linear combinations of singlet, triplet, and quintet spin-adapted excitation operators.

The closed-form expression in terms of spinorbital operators for the second-order Trotter-Suzuki decomposition of $\exp(\theta A_{PP}^{QR}/\sqrt{2})$ is given by
\begin{equation}\label{seq:trot2_closed}
	\begin{split}
		&e^{\frac{\theta}{2\sqrt{2}} A_{\Pu\Pd}^{\Qu\Rd}} e^{-\frac{\theta}{\sqrt{2}} A_{\Pu\Pd}^{\Qd\Ru}} e^{\frac{\theta}{2\sqrt{2}} A_{\Pu\Pd}^{\Qu\Rd}} = I\\
		& +\sin(\frac{\theta}{\sqrt{2}}) \left( A_{\Pu\Pd}^{\Qu\Rd} - A_{\Pu\Pd}^{\Qd\Ru} \right)\\
		&+ \left[\cos{\left(\frac{\theta}{\sqrt{2}} \right)} - 1\right] \left[\bar{n}_{P\uparrow P\downarrow} \left( n_{Q\downarrow R\uparrow} + n_{Q\uparrow R\downarrow}\right) + \left(\bar{n}_{Q\downarrow R\uparrow} + \bar{n}_{Q\uparrow R\downarrow} \right) n_{P\uparrow P\downarrow}\right]\\
		&+ \sin^2\left(\frac{\theta}{2\sqrt{2}}\right)\left[ 1 - \cos(\frac{\theta}{\sqrt{2}}) \right] \left(\bar{n}_{P\uparrow P\downarrow} n_{Q\uparrow Q\downarrow R\uparrow R\downarrow} + \bar{n}_{Q\uparrow Q\downarrow R\uparrow R\downarrow} n_{P\uparrow P\downarrow}\right)\\
		&+ \sin{\left(\frac{\theta}{2\sqrt{2}} \right)} \sin{\left(\frac{\theta}{\sqrt{2}} \right)} H^{Q\downarrow R\uparrow}_{Q\uparrow R\downarrow} \left(\bar{n}_{P\uparrow} \bar{n}_{P\downarrow} + n_{P\uparrow} n_{P\downarrow}\right)\\
		&+ \frac{1}{2}\sin^{2}{\left(\frac{\theta}{2\sqrt{2}} \right)} \left[1 - \cos{\left(\frac{\theta}{\sqrt{2}} \right)} \right] \left(\bar{n}_{\Pu\Pd} + n_{\Pu\Pd}\right) \left(\bar{n}_{\Qu\Rd} n_{\Qd\Ru} + \bar{n}_{\Qd\Ru} n_{\Qu\Rd}\right)\\
		&+ \sin\left(\frac{\theta}{\sqrt{2}} \right) \left[1 - \cos{\left(\frac{\theta}{2\sqrt{2}} \right)} \right] A_{\Pu\Pd}^{\Qd\Ru} \left(\bar{n}_{Q\uparrow R\downarrow} + n_{Q\uparrow R\downarrow}\right)\\
		&- \sin\left( \frac{\theta}{2\sqrt{2}}\right) \cos\left( \frac{\theta}{2\sqrt{2}}\right) \left[ 1 - \cos\left(\frac{\theta}{\sqrt{2}} \right) \right] A_{\Pu\Pd}^{\Qu\Rd} \left(\bar{n}_{Q\downarrow R\uparrow} + n_{Q\downarrow R\uparrow}\right)\\
		&- \frac{1}{2}\sin^{2}{\left(\frac{\theta}{2\sqrt{2}} \right)} \left[1 - \cos{\left(\frac{\theta}{\sqrt{2}} \right)} \right] \left(\bar{n}_{\Pu\Pd} - n_{\Pu\Pd}\right) \left( \bar{n}_{\Qu\Rd} n_{\Qd\Ru} - \bar{n}_{\Qd\Ru} n_{\Qu\Rd}\right).
		\end{split}
\end{equation}
By comparing \cref{seq:trot2_closed} with the closed-form expression for the first-order Trotter-Suzuki decomposition, [\cref{eq:trot1_closed} in the main text], we note that the second-order approximation captures additional terms.
As above, the terms can be grouped into singlet spin-adapted and not singlet spin-adapted.

The closed-form expression in terms of spinorbital operators for the fourth-order Trotter-Suzuki decomposition of $\exp(\theta A_{PP}^{QR}/\sqrt{2})$ is too lengthy to be presented here, but can be accessed via the Jupyter Notebook that forms part of this \sm.

\section{Derivations of Useful Expressions used in Periodicity Exploration}\label{sec:periodicity}

Let us assume that $e^{\theta A_{PP}^{QR}}$ is a 
periodic function of $\theta$ with period $T$. Then,
\begin{equation}\label{periodicity}
	\begin{split}
		e^{(\theta+T) A_{PP}^{QR}} &= e^{\theta A_{PP}^{QR}} \Rightarrow \\
		e^{\theta A_{PP}^{QR}} e^{T A_{PP}^{QR}} &= e^{\theta A_{PP}^{QR}} \Rightarrow \\
		e^{T A_{PP}^{QR}} &= \mathbf{1}.
	\end{split}
\end{equation}
Writing $A_{PP}^{QR} = U D U^\dagger$ where $U$ is unitary and $D$ diagonal, it is 
possible to derive an analytic approach to test periodicity:
\begin{equation}
	\begin{split}
		e^{T U D U^\dagger} &= \mathbf{1} \Rightarrow \\
		U e^{TD} U^\dagger &= \mathbf{1} \Rightarrow \\
		e^{TD} &= \mathbf{1} \Rightarrow \\
		e^{iT\lambda_p} &= 1, \quad p = 1,2,\ldots,\dim(D), \\
		T\lambda_p &= 2\pi n_p, \quad p = 1,2,\ldots,\dim(D) \text{ and } n_p \in \mathbb{Z},
	\end{split}
\end{equation}
where in the last steps we used the fact that anti-Hermitian matrices have purely imaginary eigenvalues. Therefore, for 
$e^{\theta A_{PP}^{QR}}$ to be periodic, the ratio of any two non-zero eigenvalues of $A_{PP}^{QR}$ must be a 
rational 
number,
\begin{equation}
	\frac{\lambda_p}{\lambda_q} = \frac{n_p}{n_q} \in \mathbb{Q}, \quad n_q \ne 0.
\end{equation}

\section{Closed-Form Expressions for the Singlet Spin-Adapted Unitaries $\boldsymbol{\exp(\theta \presuperscript{A}{[0]}{_{PQ} ^{RS}})}$ and $\boldsymbol{\exp(\theta \presuperscript{A}{[1]}{_{PQ} ^{RS}})}$}\label{sec:sa_doubles_closed}

In this section, we provide the closed-form expressions of the more complicated singlet spin-adapted unitaries considered in this work in terms of powers of their respective generators.
The corresponding expressions in terms of spinorbital operators are too lengthy to reproduce here, but can be accessed via the Jupyter Notebook that forms part of this \sm.
The closed-form expression for the unitary generated by a singlet spin-adapted double excitation going through an intermediate singlet is given by
\begin{equation}
	\begin{split}
	e^{\theta \presuperscript{A}{[0]}{_{PQ} ^{RS}}} ={}& I\\
	&+ \left[\frac{128}{21} \sin\left(\frac{\theta}{2}\right) - \frac{8\sqrt{2}}{3} \sin\left(\frac{\theta}{\sqrt{2}}\right) + \frac{2}{3} \sin\left(\theta\right) -\frac{\sqrt{2}}{42} \sin\left(\sqrt{2}\theta\right)\right] \presuperscript{A}{[0]}{_{PQ} ^{RS}}\\
	&+ \left[-\frac{256}{21} \cos\left(\frac{\theta}{2}\right) + \frac{16}{3} \cos\left(\frac{\theta}{\sqrt{2}}\right) - \frac{2}{3} \cos\left(\theta\right) + \frac{1}{42} \cos\left(\sqrt{2}\theta\right) + \frac{15}{2}\right] {\presuperscript{A}{[0]}{_{PQ} ^{RS}}}^2\\
	&+ \left[\frac{64}{3} \sin\left(\frac{\theta}{2}\right) - \frac{44\sqrt{2}}{3} \sin\left(\frac{\theta}{\sqrt{2}}\right) + \frac{13}{3} \sin\left(\theta\right) -\frac{\sqrt{2}}{6} \sin\left(\sqrt{2}\theta\right)\right] {\presuperscript{A}{[0]}{_{PQ} ^{RS}}}^3\\
	&+ \left[-\frac{128}{3} \cos\left(\frac{\theta}{2}\right) + \frac{88}{3} \cos\left(\frac{\theta}{\sqrt{2}}\right) - \frac{13}{3} \cos\left(\theta\right) + \frac{1}{6} \cos\left(\sqrt{2}\theta\right) + \frac{35}{2}\right] {\presuperscript{A}{[0]}{_{PQ} ^{RS}}}^4\\
	&+ \left[\frac{64}{3} \sin\left(\frac{\theta}{2}\right) - \frac{52\sqrt{2}}{3} \sin\left(\frac{\theta}{\sqrt{2}}\right) + \frac{22}{3} \sin\left(\theta\right) -\frac{\sqrt{2}}{3} \sin\left(\sqrt{2}\theta\right)\right] {\presuperscript{A}{[0]}{_{PQ} ^{RS}}}^5\\
	&+ \left[-\frac{128}{3} \cos\left(\frac{\theta}{2}\right) + \frac{104}{3} \cos\left(\frac{\theta}{\sqrt{2}}\right) - \frac{22}{3} \cos\left(\theta\right) + \frac{1}{3} \cos\left(\sqrt{2}\theta\right) + 15\right] {\presuperscript{A}{[0]}{_{PQ} ^{RS}}}^6\\
	&+ \left[\frac{128}{21} \sin\left(\frac{\theta}{2}\right) - \frac{16\sqrt{2}}{3} \sin\left(\frac{\theta}{\sqrt{2}}\right) + \frac{8}{3} \sin\left(\theta\right) -\frac{4\sqrt{2}}{21} \sin\left(\sqrt{2}\theta\right)\right] {\presuperscript{A}{[0]}{_{PQ} ^{RS}}}^7\\
	&+ \left[-\frac{256}{21} \cos\left(\frac{\theta}{2}\right) + \frac{32}{3} \cos\left(\frac{\theta}{\sqrt{2}}\right) - \frac{8}{3} \cos\left(\theta\right) + \frac{4}{21} \cos\left(\sqrt{2}\theta\right) + 4\right] {\presuperscript{A}{[0]}{_{PQ} ^{RS}}}^8.
	\end{split}
\end{equation}
The closed-form expression for the unitary generated by a singlet spin-adapted double excitation going through an intermediate triplet is given by
\begin{equation}
	\resizebox{\textwidth}{!}{$
	\begin{split}
		&e^{\theta \presuperscript{A}{[1]}{_{PQ} ^{RS}}} = I\\
		&+\left[ -\frac{16 \sin \left(\frac{\sqrt{3} \theta}{2}\right)}{25 \sqrt{3}}-\frac{54}{25} \sqrt{3} \sin \left(\frac{\theta}{\sqrt{3}}\right)+\frac{8}{5} \sqrt{2} \sin \left(\frac{\theta}{\sqrt{2}}\right)+\frac{432}{115} \sqrt{3} \sin \left(\frac{\theta}{2 \sqrt{3}}\right)+\frac{\sin \left(\sqrt{2} \theta\right)}{575 \sqrt{2}} \right] \presuperscript{A}{[1]}{_{PQ} ^{RS}}\\
		&+\left[ \frac{32}{75} \cos \left(\frac{\sqrt{3} \theta}{2}\right)-\frac{16}{5} \cos \left(\frac{\theta}{\sqrt{2}}\right)+\frac{162}{25} \cos \left(\frac{\theta}{\sqrt{3}}\right)-\frac{2592}{115} \cos \left(\frac{\theta}{2 \sqrt{3}}\right)-\frac{\cos \left(\sqrt{2} \theta\right)}{1150}+\frac{113}{6}\right] {\presuperscript{A}{[1]}{_{PQ} ^{RS}}}^2\\
		&+\left[-\frac{56 \sin \left(\frac{\sqrt{3} \theta}{2}\right)}{5 \sqrt{3}}-\frac{171}{5} \sqrt{3} \sin \left(\frac{\theta}{\sqrt{3}}\right)+\frac{404}{15} \sqrt{2} \sin \left(\frac{\theta}{\sqrt{2}}\right)+\frac{2952}{115} \sqrt{3} \sin \left(\frac{\theta}{2 \sqrt{3}}\right)+\frac{11 \sin \left(\sqrt{2} \theta\right)}{345 \sqrt{2}} \right] {\presuperscript{A}{[1]}{_{PQ} ^{RS}}}^3\\
		&+\left[ \frac{112}{15} \cos \left(\frac{\sqrt{3} \theta}{2}\right)-\frac{11}{690} \cos \left(\sqrt{2} \theta\right)-\frac{808}{15} \cos \left(\frac{\theta}{\sqrt{2}}\right)+\frac{513}{5} \cos \left(\frac{\theta}{\sqrt{3}}\right)-\frac{17712}{115} \cos \left(\frac{\theta}{2 \sqrt{3}}\right)+\frac{587}{6} \right] {\presuperscript{A}{[1]}{_{PQ} ^{RS}}}^4\\
		&+\left[ -\frac{1192 \sin \left(\frac{\sqrt{3} \theta}{2}\right)}{25 \sqrt{3}}-\frac{2718}{25} \sqrt{3} \sin \left(\frac{\theta}{\sqrt{3}}\right)+\frac{133 \sqrt{2} \sin \left(\sqrt{2} \theta\right)}{1725}+\frac{308}{3} \sqrt{2} \sin \left(\frac{\theta}{\sqrt{2}}\right)+\frac{1368}{23} \sqrt{3} \sin \left(\frac{\theta}{2 \sqrt{3}}\right)\right] {\presuperscript{A}{[1]}{_{PQ} ^{RS}}}^5\\
		&+\left[ \frac{2384}{75} \cos \left(\frac{\sqrt{3} \theta}{2}\right)-\frac{616}{3} \cos \left(\frac{\theta}{\sqrt{2}}\right)+\frac{8154}{25} \cos \left(\frac{\theta}{\sqrt{3}}\right)-\frac{8208}{23} \cos \left(\frac{\theta}{2 \sqrt{3}}\right)-\frac{133 \cos \left(\sqrt{2} \theta\right)}{1725}+\frac{613}{3} \right] {\presuperscript{A}{[1]}{_{PQ} ^{RS}}}^6\\
		&+\left[ \frac{16}{115} \sqrt{2} \sin \left(\sqrt{2} \theta\right)+\frac{608}{5} \sqrt{2} \sin \left(\frac{\theta}{\sqrt{2}}\right)-\frac{112}{5} \sqrt{3} \sin \left(\frac{\sqrt{3} \theta}{2}\right)-\frac{576}{5} \sqrt{3} \sin \left(\frac{\theta}{\sqrt{3}}\right)+\frac{6192}{115} \sqrt{3} \sin \left(\frac{\theta}{2 \sqrt{3}}\right)\right] {\presuperscript{A}{[1]}{_{PQ} ^{RS}}}^7\\
		&+\left[ \frac{224}{5} \cos \left(\frac{\sqrt{3} \theta}{2}\right)-\frac{16}{115} \cos \left(\sqrt{2} \theta\right)-\frac{1216}{5} \cos \left(\frac{\theta}{\sqrt{2}}\right)+\frac{1728}{5} \cos \left(\frac{\theta}{\sqrt{3}}\right)-\frac{37152}{115} \cos \left(\frac{\theta}{2 \sqrt{3}}\right)+176 \right] {\presuperscript{A}{[1]}{_{PQ} ^{RS}}}^8\\
		&+\left[ \frac{48}{575} \sqrt{2} \sin \left(\sqrt{2} \theta\right)+\frac{192}{5} \sqrt{2} \sin \left(\frac{\theta}{\sqrt{2}}\right)-\frac{192}{25} \sqrt{3} \sin \left(\frac{\sqrt{3} \theta}{2}\right)-\frac{864}{25} \sqrt{3} \sin \left(\frac{\theta}{\sqrt{3}}\right)+\frac{1728}{115} \sqrt{3} \sin \left(\frac{\theta}{2 \sqrt{3}}\right)\right] {\presuperscript{A}{[1]}{_{PQ} ^{RS}}}^9\\
		&+\left[ \frac{384}{25} \cos \left(\frac{\sqrt{3} \theta}{2}\right)-\frac{48}{575} \cos \left(\sqrt{2} \theta\right)-\frac{384}{5} \cos \left(\frac{\theta}{\sqrt{2}}\right)+\frac{2592}{25} \cos \left(\frac{\theta}{\sqrt{3}}\right)-\frac{10368}{115} \cos \left(\frac{\theta}{2 \sqrt{3}}\right)+48 \right] {\presuperscript{A}{[1]}{_{PQ} ^{RS}}}^{10}.
	\end{split}
$}
\end{equation}

\end{document}